\documentclass[prb,showpacs,superscriptaddress,preprintnumbers,amssymb,twocolumn,amsfonts]{revtex4}
\usepackage{graphicx}
\usepackage{amssymb}
\usepackage{amsmath}
\usepackage{subfigure}
\usepackage{bm}

\begin{document}

\input epsf
\def\beq{\begin{equation}}
\def\eeq{\end{equation}}
\def\beqn{\begin{eqnarray}}
\def\eeqn{\end{eqnarray}}
\def\etal{\emph{et al.}}
\def\ket#1{\vert #1 \rangle}
\def\bra#1{\langle #1 \vert}
\def\ev#1{\langle #1 \rangle}
\def\ip#1#2{\langle #1 \vert #2 \rangle}
\def\me#1#2#3{\langle #1 \vert #2 \vert #3 \rangle}
\renewcommand{\bf}{\mathbf}

\title{Magnetic phase diagram of a spin-1 condensate in two dimensions with dipole interaction}

\author{Jonas A.~Kj\"{a}ll}
\affiliation{Department of Physics, University of California,
Berkeley, CA 94720}
\author{Andrew M.~Essin}
\affiliation{Department of Physics, University of California,
Berkeley, CA 94720}
\author{Joel~E.~Moore}
\affiliation{Department of Physics, University of California,
Berkeley, CA 94720} \affiliation{Materials Sciences Division,
Lawrence Berkeley National Laboratory, Berkeley, CA 94720}
\date{\today}

\begin{abstract}
Several new features arise in the ground-state phase diagram of a spin-1 condensate trapped in an optical trap when the magnetic dipole interaction between the atoms is taken into account along with confinement and spin precession.  The boundaries between the regions of ferromagnetic and polar phases move as the dipole strength is varied and the ferromagnetic phases can be modulated.  The magnetization of the ferromagnetic phase perpendicular to the field becomes modulated as a helix winding around the magnetic field direction, with a wavelength inversely proportional to the dipole strength.  This modulation should be observable for current experimental parameters in $^{87}$Rb.  Hence the much-sought supersolid state, with broken continuous translation invariance in one direction and broken global $U(1)$ invariance, occurs generically as a metastable state in this system as a result of dipole interaction.  The ferromagnetic state parallel to the applied magnetic field becomes striped in a finite system at strong dipolar coupling.
\end{abstract}
\maketitle

\section{Introduction}
Bose condensates of atoms with nonzero total spin $F \geq 1$ show various phases combining magnetic and superfluid order.  When the magnetic symmetry is broken spontaneously, as can occur when the atoms are confined in a spin-independent optical trap, condensates are classified as ``polar'' (for antiferromagnetic interactions) or ``ferromagnetic''.  Most theoretical studies of these spinor Bose condensates neglect the long-range interaction between atomic magnetic moments, and this neglect is justified for many experimental conditions.  However, recent experiments~\cite{Sadler,Veng1,Veng2} investigating ordering in a nearly two-dimensional condensate have shown complex magnetic behavior in the ferromagnetically interacting $F=1$ spinor Bose gas of $^{87}$Rb.

The most surprising feature of these experiments, which image the spin distribution in real space, is a long-lived phase that appears to have the broken global $U(1)$ invariance of a superfluid along with possible breaking of the continuous translational symmetry in one or two directions, i.e., with stripe-like or checkerboard-like order.  A possible supersolid phase has recently also been suggested in the superfluid of $^4$He.~\cite{KimChan}  Many theoretical papers have been written about the properties of $^4$He and whether a supersolid phase can exist in the absence of disorder.  Only recently have theoretical studies been done to explain the observed supersolid-like behavior in a $^{87}$Rb spinor condensate.~\cite{CherngDemler}  The earlier studies of $^{87}$Rb concentrated on magnetic properties arising from the weak spin-dependent local interaction and the quadratic Zeeman shift.
More recent experiments~\cite{Veng1,Veng2} indicates that the long range dipole interaction also plays an important role in the formation of the magnetic phases in spatially large systems and with this addition a supersolid state might be possible.\\  
\\
Most previous studies of this system concern dynamical properties: the leading instability when the Hamiltonian is changed to favor ferromagnetic order can be stripe-like or checkerboard-like depending on parameters.~\cite{Lamacraft,CherngDemler,sau} 
In this paper, our goal is to determine the static ground-state phase diagram.  We start from the phases that are well established at low temperatures~\cite{Ho,ohmi98,Mukerjee1,podolskychandra} for a spin-1 gas with no dipole interaction and quadratic Zeeman effect. (Low temperatures mean below the superfluid and magnetic transitions, where all the studies in this paper will take place.) We then add the dipole interaction to see how it changes the phases as well as the location of the boundary between them. We do this in a quasi-two-dimensional geometry as in the experiments.~\cite{Sadler, Veng1,Veng2}  We investigate both an infinite and a finite square planar geometry.  After observing the formation of two kinds of stripe order in a Monte Carlo simulation, we developed an analytical approach to explain the results, based upon smallness of the dipolar coupling at short distances.  That analytical approach is presented first in order to prepare the groundwork for the simulation results.


We show that all boundaries in the phase diagram, except between the two polar phases, are moved when the dipole interaction is added, some in a non-intuitive way.  The magnetic dipole interaction prefers a ferromagnetic state, but the confinement makes a ferromagnetic state out of the plane energetically unfavorable.  Moreover, the spin precession make the in-plane perpendicular ferromagnetic state unfavorable, since the spin rotates out of the plane.  Both ferromagnetic phases can get modulated in one direction.  The phase parallel to the external fields needs a strong dipole interaction or a system much wider than its length to become modulated.  This modulation appears as fully magnetized stripes with sharp domain walls between them.  The phase perpendicular to the external fields gets modulated, from the very lowest dipole strengths, into a helical configuration around the field.  The wavelength of the helix is inversely proportional to the dipole strength. For $^{87}$Rb the wavelength is $\sim 80 \mu \mathrm{m}$ and should be observable in experiments.

The outline of this paper is as follows.  In the following section, we review the basic physics of spinor condensates without the dipole interaction.  In Section III we introduce the dipole interaction and put it into a form that is convenient for numerical simulations.  Section IV presents analytical results in the limit of weak dipole interaction, and Section V contains the results of our Monte Carlo simulations of the problem.  The final section summarizes the relationship between our results and those of other theoretical papers and suggests how future experiments could be designed to observe clearly the metastable supersolid phase found in our simulations.

\section{Review of spinor condensate without magnetic dipole interaction}
A Bose-Einstein condensate of spin $F=1$ atoms
is described by a three-component complex order parameter
\begin{equation}\label{eq:op}
\Psi(\bf{x})=\sqrt{n_{3D}(\bf{x})} \psi(\bf{x})
=\sqrt{n_{3D}(\bf{x})}\left(
\begin{array}{c}
\psi_{+1}(\bf{x})\\
\psi_0(\bf{x})\\
\psi_{-1}(\bf{x})
\end{array}
\right),
\end{equation}
where the spinor $\psi(\bf{x})$ is normalized as $\psi^\dagger \psi =1$
and the subscripts label the spin eigenvalue with respect to an
arbitrarily chosen quantization direction.  In the absence of
external fields and neglecting the dipole interaction, the Hamiltonian 
governing the condensate is\cite{Ho,ohmi98}
\begin{equation}\label{ham}
H_0=\int d^3x  \left[
\frac{\hslash^2}{2m} \left\vert \nabla\Psi \right\vert^2
+\frac{c_0}{2} n_{3D}^2
+\frac{c_2}{2} n_{3D}^2 M^2 \right],
\end{equation}
where $m$ is the atomic mass, 
$\bf{M}(\bf{x}) = \psi^\dag(\bf{x}) \bf{F} \psi(\bf{x})$ is the
dimensionless magnetization ($|\bf{M}|\leq 1$), and $\{F^i\}$ are 
the three generators of $SU(2)$ in the spin-1 representation
\begin{align}
\label{s1m}
\nonumber F^x=\frac{1}{\sqrt{2}}
\begin{pmatrix}
0& 1& 0\\
1& 0& 1\\
0& 1& 0
\end{pmatrix}&, \quad
F^y=\frac{1}{\sqrt{2}}
\begin{pmatrix}
0& -i& 0\\
i& 0& -i\\
0& i& 0
\end{pmatrix}, \\
F^z=&
\begin{pmatrix}
1& 0& 0\\
0& 0& 0\\
0& 0& -1
\end{pmatrix}.
\end{align}
The first term in the Hamiltonian is the kinetic energy for bosons with 
mass $m$. The next two terms are the spin-independent and 
spin-dependent contact interactions, respectively. The coefficients are 
given by $c_0=(4\pi\hslash^2/3m)(2a_2+a_0)$ and 
$c_2=(4\pi\hslash^2/3m)(a_2-a_0)$, with $\{a_0,a_2\}$ the s-wave 
scattering lengths in the channel with total angular momentum $\{0,2\}$.

When $c_2<0$ (``ferromagnetic'') it is energetically favorable for this 
system to magnetize, $M\neq 0$, while $c_2>0$ favors a ``polar'' state
with $M=0$.  The scattering lengths for $^{87}$Rb are $a_0=101.8a_B$ and 
$a_2=100.4a_B$,~\cite{Kempen} where $a_B$ 
is the Bohr radius, so $c_2$ is negative 
and its condensate will be ferromagnetic in the absence of external
fields (still neglecting the dipole interaction). However, the condensate of
$^{23}$Na will be in a polar state.~\cite{Ho}

The external fields normally applied to a spinor condensate consist of an optical trap and a uniform magnetic field described by the following addition to the Hamiltonian
\beq
\label{hamef}
H_{ef}=\int \!d^3x \left[ U 
+q\psi^\dag(\hat{\bf{B}} \cdot \textbf{F})^2 \psi
\right]n_{3D},
\eeq
The trapping potential $U(\bf{x})$ confines the condensate spatially;
for our purposes, its main effect will be to produce a 
quasi-two-dimensional geometry.  The quadratic Zeeman shift $q$ can be tuned 
independently of $\bf{B}$ with microwave radiation, $q=q^B+q^{EM}$.~\cite{Kater}  We take the two
sources as coaxial along $\hat{\bf{z}}$, so we can use Eq.~\eqref{hamef}.  This is also the axis we quantize the spinor along.  The magnetic field also creates a linear Zeeman term $\bf{B}\cdot\int \!d^3x \,n_{3D}\bm{\mu}$, that favors an uniformly 
magnetized condensate.  However,  experiments on $^{87}$Rb have not observed any tendency toward such relaxation over
the accessible time scales of several seconds,~\cite{Kater} making the longitudinal component of magnetization conserved.  (This assumption does not apply in condensates of higher spin, such as chromium.~\cite{griesmaier05})  Normally, this component is chosen to vanish initially and can hence be ignored for the purpose of energetics.  However, the magnetic field also causes Larmor precession of the magnetization perpendicular to it.  This is an important effect that needs to be taken into account as it modifies the nature of the magnetic interaction on time scales longer than the precession time.

\begin{figure}[htbp]
  \begin{center}
    \subfigure{\includegraphics[width=30mm]{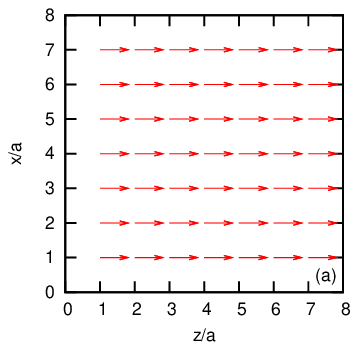}}
        \subfigure{\includegraphics[width=30mm]{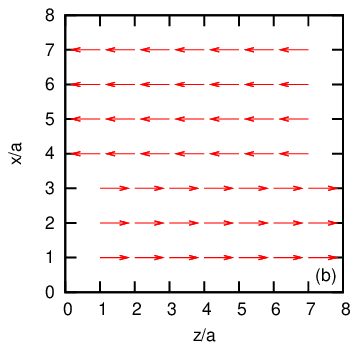}}\\
            \subfigure{\includegraphics[width=30mm]{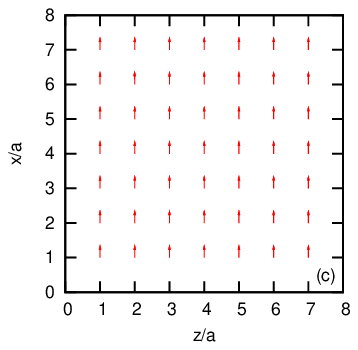}}
                \subfigure{\includegraphics[width=30mm]{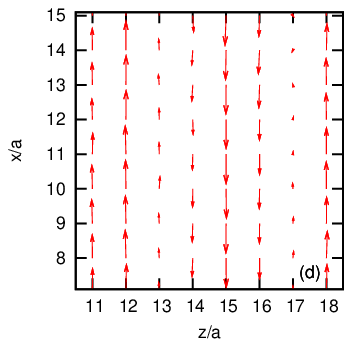}}
    \caption{(Color online) Examples of possible spin configurations in the plane. The external fields are along the horizontal axis, $M_z(x,z)$ is plotted on the horizontal axis, $M_x(x,z)$ is on the vertical for every plaquette and $M_y(x,z)$ is not shown. (a) Uniform fully magnetized $F_{\|}$, (b) striped fully magnetized $F_{\|}$, (c) uniform partly magnetized $F_{\perp}/P_{\|}$ state, (d) helical fully magnetized $F_{\perp}$ .}
    \label{phases}
  \end{center}
\end{figure}
\begin{figure}[htbp]
  \begin{center}
    \includegraphics[width=70mm]{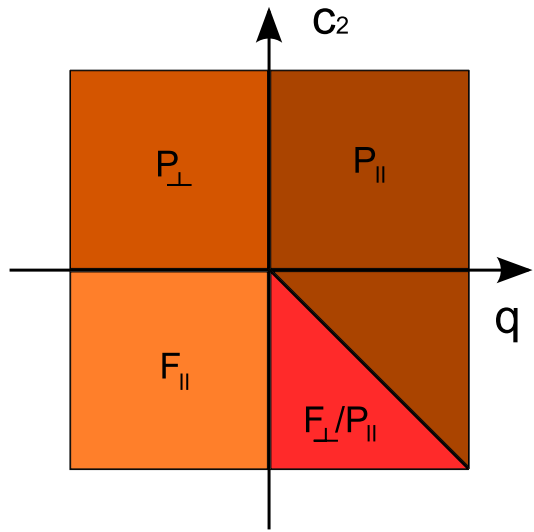}
    \caption{(Color online) Ground state phase diagram of a spin-1 
		condensate 
		without dipolar interaction; from Mukerjee \etal.~\cite{mukerjee}}
    \label{c2qpd}
  \end{center}
\end{figure}

The spin state of the condensate is governed by 
the parameters $c_2$ and $q$, as in 
Fig.~\ref{c2qpd}.~\cite{mukerjee}  There are two different kinds of polar states ($c_2>0$), one that minimizes $\ev{(F^z)^2}=0$ and one that maximizes $\ev{(F^z)^2}=1$ the impact of the quadratic Zeeman term.  Respectively,
\begin{equation}\label{polarpsi}
 \psi^P_\parallel(\phi) 
 = e^{i\phi}\begin{pmatrix} 0\\1\\0 \end{pmatrix}, \quad
 \psi^P_\perp(\phi,\theta) 
 = \frac{e^{i\phi}}{\sqrt{2}}
 \begin{pmatrix} -e^{-i\theta}\\0\\e^{i\theta} \end{pmatrix}.
\end{equation}
Consequently, does the phase $P_\parallel$, with order-parameter manifold $U(1)$,
appears at $q>0$, while
the phase $P_\perp$ appears when $q<0$.  Note that the range
of $\theta$ is only $[0,\pi)$, or alternatively that the 
order-parameter
manifold for this phase is $U(1)\times U(1)/Z_2$.~\cite{zhou01}  When 
$c_2<0$ and $q<0$, both energies are minimized by 
ferromagnetic states
\begin{equation}
\label{spfpa}
 \psi^F_{\parallel,\uparrow}(\phi)
 = e^{i\phi}\begin{pmatrix} 1\\0\\0 \end{pmatrix}, \quad
 \psi^F_{\parallel,\downarrow}(\phi)
 = e^{i\phi}\begin{pmatrix} 0\\0\\1 \end{pmatrix},
\end{equation}
giving a manifold $U(1)\times Z_2$ (recall that we exclude the linear
Zeeman energy from energetic considerations), see Fig.~\ref{phases}.  
In the final quadrant of the phase diagram, however, no
ferromagnetic state minimizes the quadratic Zeeman energy.  The smallest impact of a ferromagnetic state on the quadratic Zeeman term is $\ev{(F^z)^2}=1/2$ for 
\begin{equation}\label{Fperppsi}
 \psi^F_\perp(\phi,\xi)
 = \frac{e^{i\phi}}{2}
 \begin{pmatrix} e^{-i\xi}\\ \sqrt{2}\\e^{i\xi} 
 \end{pmatrix},
\end{equation}
Consequently, for $q<q_c=2|c_2|n_{3D}$ the state will be a linear
combination of $\psi^P_\parallel$ and $\psi^F_\perp$
with magnetization $M_x + i M_y  
= \sqrt{1-(q/q_c)^2}\,e^{i\xi}$ and manifold 
$U(1)\times U(1)$, see Fig.~\ref{phases}.  Above $q_c$, the state will be the pure polar state $P_\parallel$. 

Typical experimental values for 
$^{87}$Rb~\cite{Sadler,Veng1,Veng2} include a peak density 
of $n_0=2.5 \times 10^{14}\,\mathrm{cm}^{-3}$, giving the 
interaction strengths $c_0n_0=1.9\,\mathrm{kHz}$ and 
$c_2n_0=-9\,\mathrm{Hz}$, while $q^B\approx1.6\mathrm{Hz}$ and $q^{EM}$ can be tuned from 
roughly $-50\,\mathrm{Hz}$ to $50\,\mathrm{Hz}$ and is normally taken coaxial to $q^B$.~\cite{Kater}

\subsection{Confinement}\label{confinement}
The optical trap in the experiment makes the gas effectively two dimensional, with a Thomas-Fermi radius $r_{TF} \approx1.5\mu m$ along the direction of tightest confinement.~\cite{Sadler,Veng1,Veng2}  Since this is smaller than the spin healing length $\xi=\sqrt{\hbar^2/(2m|c_2|n_{3D})} \approx 2.5 \mu m$, we take the confinement to be along the $\hat{y}$ direction and treat the gas as frozen along this direction; that is, we take
\beq
\Psi(\bf{x}) = \sqrt{ n_{2D}(x,z) \rho(y) } \, \psi(x,z),
\eeq
where we assume $\int \!dy \,\rho(y)=1$.
In the following we will consider one of two profiles $\rho(y)$ as convenient, a boxcar profile and a Gaussian,
 \begin{equation}
  \label{eq:3to2u}
  \rho_1(y) = \frac{1}{K}\Theta(K/2-y)\Theta(K/2+y), \;
  \rho_2(y) = \frac{1}{\sigma_y} 
\sqrt{\frac{2}{\pi}} e^{-\frac{2y^2}{\sigma_y^2}},
\end{equation} 
where $\Theta(x)$ is the Heaviside step function.  We introduce a common notation for the condensate thickness $T$ and a 3-dimensional density $\bar{n}_{3D}(x,z)$ without y-dependence
\beq
\label{com}
\frac{1}{T}= \ev{\rho} = \int \!dy\, \rho(y)^2 , \hspace{3mm} \bar{n}_{3D}(x,z)=\frac{n_{2D}(x,z)}{T}
\eeq
for the boxcar profile and for the gaussian profile, to be able to treat both profiles simultaneously in section \ref{sec:Res}.  In most of our analysis these densities are also independent of $(x,z)$, except where we use a nonzero trapping potential $U(x,z)$ in the plane.

\subsection{Precession}
Atoms with magnetic moment $\bm{\mu}_{\perp}=g_F \mu_B\textbf{M}_{\perp}$ perpendicular to the field precess at
frequency $|\gamma| B_0 = |g_F| \mu_B B_0$ around the fields.  As usual, $\mu_B$ is the Bohr 
magneton and $g_F$ is Lande's g-factor.  For $^{87}$Rb, 
$g_F = -1/2$ and a field of $B_0 = 150\,\mathrm{mG}$
produces a Larmor precession at 110 kHz, a scale orders of magnitude larger than 
the contact interactions or the quadratic Zeeman energy.

The Hamiltonian considered so far is invariant under the spin rotation
\begin{equation}
  \label{Rot}
\psi _k(x,z) \rightarrow U_{kl}(t)\psi_l(x,z),\quad
 U(t)=e^{-i\gamma B_0\hat{\bf{B}} \cdot \textbf{F} t}
\end{equation}
and is hence unaffected by the rapid Larmor precession. Therefore, 
adding precession does not affect the phase diagram in the problem with only local interactions.~\cite{Ho,ohmi98}  However when we include the dipole
interaction in the next section, both confinement and spin precession 
become important.
      
\section{Magnetic dipole interaction}
The interactions considered thus far for a spin-1 condensate are all local. However, the
moments $\bm{\mu}$ will interact through the long-ranged dipole
interaction. This is weak for $^{87}$Rb relative to most other energies
in the system, but since it is long ranged it will have an important
impact on the magnetic phases. The initial studies of the spin-1
condensate ignored this term,~\cite{Ho,mukerjee} but some recent works
have included it along with the effects of quasi-two-dimensional
confinement and rapid Larmor precession.~\cite{CherngDemler,Lamacraft}
Among other results, it was shown that dipolar interaction renders the
Larmor precession unstable,~\cite{Lamacraft} and we return to this point in
the concluding section.  Until then we follow previous authors and assume
that this instability has significant effects only at late times, and
so neglect it. Cherng and Demler examined the instability spectrum of a
uniform ferromagnetic state within a mean field and collective mode
analysis.  We will use the same physical model but instead look at the ground
state phase diagram and consider a wider range of parameters $c_2$, $q$, and
$c_d$ (see Eq.~\eqref{Tham} below) with analytical and Monte Carlo
calculations.

The total Hamiltonian we work with is
\beq
\label{Tham}
H=H_0+H_{ef}+H_{dip}
\eeq
where
\begin{multline}
 \label{dipham}
H_{dip} = \frac{c_d}{2}\! \int\! d^3xd^3x' n_{3D}(\bf{x}) 
M_i(\bf{x})n_{3D}(\bf{x}')M_j(\bf{x}') \\
 \times \left[\nabla_i\nabla'_j\frac{1}{|\bf{x}-\bf{x}'|} - 
 \frac{4\pi}{3} \delta_{ij}\delta^{(3)}(\bf{x}-\bf{x}')\right].
\end{multline}
This is the same as the more usual expression with
$(\delta_{ij}-3\hat{r}_i\hat{r}_j)/r^3$, but split it into a part that
is positive-(semi-)definite and a part that simply shifts the
parameter $c_2 \rightarrow c_2 - 4\pi c_d/3$ (see the beginning of Appendix \ref{app:dipole} for a fuller discussion of the magnetic dipole term).  Indeed, with two integrations by parts the first term
becomes the Coulomb interaction for a charge density 
$\bm{\nabla}\cdot (n_{3D}\bf{M})$.  We will typically mean just this term 
when referring to ``the dipole interaction,'' since it is the difficult 
part.
The strength of the dipole term is given by $c_d=\mu_0g_F^2\mu_B^2/4\pi$, 
where $\mu_0$ is the vacuum permeability, giving a value of 
$c_d n_0=0.8\,\mathrm{Hz}$ for $^{87}$Rb.  
The effect of confinement is less trivial for this term then for the
others, and transforming to a rotating frame is also nontrivial since
the interaction couples spin directions to spatial directions.  See
Appendix~\ref{app:dipole} for a full treatment of these effects.  In the following section, we discuss how the dipole interaction is expected to modify the phase diagram when it is sufficiently weak that the $H_{dip}=0$ ground states can be used as a starting point.


\section{Analytical Results}\label{sec:Res}
Adding the dipole interaction Eq.~(\ref{dipham}) will change the phase diagram Fig.~\ref{c2qpd}. The term that looks like the spin dependent interaction will just move the whole phasediagram up along $c_2$ with $\frac{4\pi c_d}{3}$.  The energy from the Coulomb part of the dipole interaction is always positive, hence this parts prefers a polar state with zero magnetization  $M=0$. 
Consequently, regions of Fig.~\ref{c2qpd} with polar states above $c_2=\frac{4\pi c_d}{3}$ will not 
change if we add the dipole-dipole coupling.  However, the rest of the 
phase diagram may be affected and the phase boundaries will depend on $c_d$, as we now discuss in 
some detail.

\subsection{Weak dipole interaction}
Adding a weak dipole term (weak compared to the kinetic energy term) will only change the phase diagram slightly.  We start out by ignoring any new phases and investigate how a weak dipole interaction will move the boundaries between the existing phases.  The three magnetic terms in the Hamiltonian are the spin-dependent 
contact interaction, the quadratic Zeeman and the dipole term.  By 
comparing the energy contributions from these three for simple 
\emph{Ans\"{a}tze} we can locate the boundaries between different 
minima, in a system with $L$ the extent along $z$ and $W$ the extent along $x$.

The polar phases are, of course, the simplest (see 
Eq.~\eqref{polarpsi})
\begin{equation}
E^{P_{\|}} = 0, \quad E^{P_{\perp}} = q \bar{n}_{3D}LWT.
\end{equation}
Consider next the phase $F_{\|}$, which appeared at $q,c_2<0$ in the 
system without dipolar energy.  The effective charge density for such
a state describes two quasi 1-dimensional lines of charge located at
$\pm L/2$, of length $W$.  The self-energy of such lines of charge is
given by $2c_d (\bar{n}_{3D} M)^2 WT^2\ln{W/T}$, see Appendix \ref{app:Dipener}, to leading order.  The other two terms are easily kept exact. Keeping terms of order $A^2$ and $A\ln{A}$ where $A=L,W$, (see 
Eq.~(\ref{spfpa})) 
\begin{equation}
E^{F_{\|}} = \frac{\tilde{c}_2}{2} \bar{n}_{3D}^2 LWT + q \bar{n}_{3D}LWT, 
+ 2c_d \bar{n}_{3D}^2 WT^2\ln W/T
\end{equation}
with $\tilde{c}_2 = c_2 - 4\pi c_d/3$.

\begin{figure}[htbp]
  \begin{center}
    \includegraphics[width=70mm]{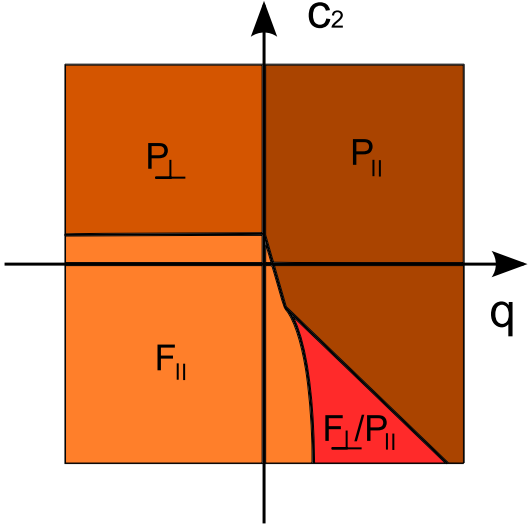}
    \caption{(Color online) Ground state phase diagram for a spin-1 condensate with dipole interaction and external fields, that introduces a quadratic Zeeman term and rapid spin precession. Both polar and ferromagnetic phases appear, perpendicular as well as parallel to the field. }
    \label{cdpd}
  \end{center}
\end{figure}

The transition in the left half-plane betwen the states $F_{\|}$ and $P_{\perp}$, see Fig.~\ref{cdpd}, will hence be moved up from $c_2=0$ for a system without dipole interaction to
\begin{equation}
  \label{FtP}
c_{2c} \equiv 4c_d \left(\frac{\pi}{3}-\epsilon_L\right),
\end{equation}
where $\epsilon_L=\frac{\ln{W/T}}{L/T}$ will vanish in the large-system limit.

The region of the phase diagram with $q>0$ and $c_2<0$, is the most
interesting, due to the rapid precession of the perpendicular magnetization about the magnetic field, and the high dipolar energy cost of spins pointing out of the plane.  Consequently, the region of $F_{\perp}/P_{\|}$ in the phase diagram will shrink and the regions of $P_{\|}$ and $F_{\|}$ grow, with the latter extending to positive values of $q$.  For a uniform condensate with spins out of the plane, the Coulomb energy is equivalent to that of a 
parallel-plate capacitor, giving an energy $2\pi c_d (\bar{n}_{3} M)^2 T^2LW/T$
to leading order, \emph{i.e.}, neglecting fringing fields, see Appendix \ref{app:Dipener}. 

Because of
the precession, the spins will effectively average the out-of-plane and in-plane interaction energies with equal weights. Consequently, the dipole energy for magnetization perpendicular to the external fields is $c_d (\bar{n}_{3} M)^2(\pi W + T\ln{L/T})LT$. To find the energy for the $F_{\perp}/P_{\|}$ state, we first have to find $M$, since this state is not completly magnetized.  Consider a spinor $\psi^T=(a,b,a)$ with $a=\sqrt{(1-b^2)/2}$ ($1/\sqrt{2}<b<1$), which represents a 
superposition of $\psi^P_\|$ and $\psi^F_\perp$ (see
Eqs.~\eqref{polarpsi} and \eqref{Fperppsi}).  Its magnetization is 
$M_x = 2b\sqrt{1-b^2}$.  Putting it all together,
\begin{multline} \label{EF}
E^{F_{\perp}/P_{\|}}
= 4b^2(1-b^2)(\frac{\tilde{c}_2}{2}W+ c_d(\pi W + T\ln{W/T}))\bar{n}_{3D}LT \\ 
+ q \bar{n}_{3D}(1-b^2) LWT.	
\end{multline}
The energy for this state is minimized at
\begin{equation}  \label{min} 
b^2 = \frac{1}{2} \left( 1 + \frac{q}{q_c} \right).
\end{equation}
As the notation suggests, the transition between the phases $P_{\|}$
and $F_{\perp}/P_{\|}$ occurs at $q=q_c$, where $E^{F_{\perp}/P_{\|}}=E^{P_{\|}}=0$ and $M=0$
\begin{equation}
\label{qc}
q_c \equiv 2|c_2|\bar{n}_{3D} - 4c_d\bar{n}_{3D}\left(\frac{\pi}{3} +\epsilon_W \right), 
\end{equation}
where $\epsilon_W=\frac{\ln{L/T}}{W/T}$ will vanish in the large system limit.
As can be seen in Eqs.~(\ref{min}), (\ref{qc}) and (\ref{mf}), the value of the magnetization and hence the order parameter for the $F_{\perp}/P_{\|}$ state decreases continuously and is zero at the phase transition to the $P_{\|}$ state.
\begin{equation}
\label{mf}
M_0=|\langle F_{\bot}/P_{\|}\rangle |=\sqrt{1-(q/q_c)^2}
\end{equation}
This is exactly the same equation as for a system without dipole interaction, except that $q_c$ now is given by Eq.~(\ref{qc}). 

Plugging the form for $b$, Eq.~(\ref{min}) and (\ref{qc}), back in also allow us to locate the 
transition between $F_{\perp}/P_{\|}$ and $F_{\|}$, where
$E^{F_{\perp}/P_{\|}}=E^{F_{\|}}$, which will occur at
\beq\label{qc2}
q_{c2} \equiv \sqrt{q_c\left( 2|c_2|\bar{n}_{3D}+8c_d\bar{n}_{3D}\left(\frac{\pi}{3} - 2\epsilon_L \right) \right)} - q_c.
\eeq
Finally, the transition between $F_{\|}$ and $P_{\|}$ will
take place when $E^{F_{\|}} = E^{P_{\|}}=0$, at
\beq
\label{qc3}
q_{c3} \equiv \frac{|c_2|\bar{n}_{3D}}{2} + 2c_d\bar{n}_{3D}\left(\frac{\pi}{3}-\epsilon_L\right).
\eeq
The three transition lines ($q_c$, $q_{c2}$ and $q_{c3}$) separating the three phases in the lower right quadrant meet at the point 
\beq
(q, c_2) = 4 c_d 
\left(\bar{n}_{3D}\left( \frac{\pi}{3} + \epsilon_W - 2\epsilon_L\right), 
\left( \frac{\pi}{3} + 2\epsilon_W - \epsilon_L\right)\right).
\eeq
To finish the phase diagram, we see that the transition line in Eq.~(\ref{qc3}), that separates $F_{\|}$ and $P_{\|}$, can be extended to the region $q,c_2>0$, with the substitution $|c_2| \rightarrow -c_2$ and that it will intersect with the transition line in Eq.~(\ref{FtP}) at the point $(q,c_2)=(0,c_{2c})$. 

\subsection{Magnetization textures}
The dipolar energy favors spatially modulated ferromagnetic states,
which screen the long-ranged interaction, over uniform states.  Consider
the state $F_{\|}$.  We can adapt a classic argument of Kittel 
concerning the formation of magnetic domains to the present 
quasi-two-dimensional geometry.~\cite{Kittel} The boundary energy 
$ 2c_d \bar{n}_{3D}^2 WT^2 \ln{W/T}$
from before will become $ 2c_d \bar{n}_{3D}^2 WT^2 \ln{d/T}$ if the uniform state 
breaks up into Ising-like domains of width $d$ and length $L$ that alternate between 
$M_z=1$ and
$M_z=-1$, keeping the total magnetization $M_0=1$ everywhere, see Fig.~\ref{phases}.   There 
will be a cost in kinetic energy at the domain walls, and the competition
between these two effects sets the domain size.  

We can estimate an upper
bound for the domain wall energy by assuming its width is the 
spin-healing length $\xi_S$.  The energy will scale with the area of 
the wall $\sim LT$, and the surface density will be 
$\sigma_W \sim \hslash^2 \bar{n}_{3D}/2 m \xi_S$.  With the number of domains
given by $W/d$, the energy is
\beq
\label{dom}
E = \sigma_W \frac{LWT}{d} +  2c_d \bar{n}_{3D}^2 WT^2 \ln{d/T},
\eeq
which gives
\beq
 d^\| = \frac{\sigma_W}{c_d \bar{n}_{3D}^2T} L.
\eeq
The resulting domains have a width proportional to the length
of the system, and are very large when the dipolar coupling is
weak.  In $^{87}$Rb with the experimental parameters given in 
section~\ref{confinement}, 
$\sigma_W \sim 10^4\,\mathrm{Hz\,\mu m^{-1}}$
and $d^\| \sim 20 L$, which could be difficult to achieve 
experimentally.

For a rectangular sample ($L>W$) in the $F_{\|}$ state, with a constraint of zero total longitudinal magnetization ($\int d{\bf{x}}n_{3D}(\bf{x})M^z(\bf{x}))=0$), it can be more energetically favorable to split up into two domains perpendicular to the field.  The energy for this configuration is $E = 2\sigma_W WT +  3c_d \bar{n}_{3D}^2 WT^2 \ln{W/T}$  to leading order and if this is lower than the energy in Eq.~\eqref{qc2} it will occur.  However, this is only due to the constraint; a domain-free configuration has lower energy and a configuration with several domain walls perpendicular to the field will not be favorable for any values in the phase diagram.

\begin{figure}[htbp]
  \begin{center}
    \includegraphics[width=85mm]{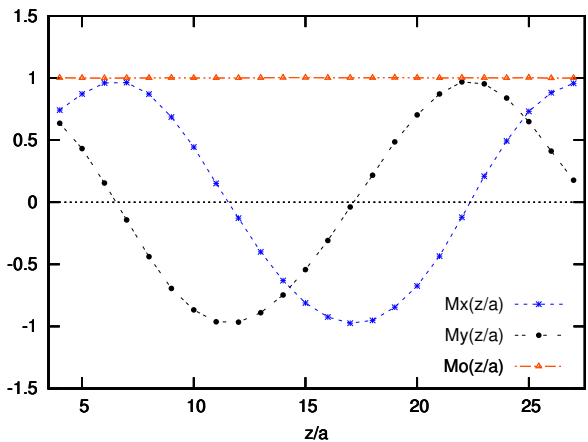}
    \caption{(Color online) Transverse magnetization as a function of z, from numerical simulation. Orange: magnitude of total magnetization $M_0$, blue: transverse magnetization in plane $M_x$ and black: transverse magnetization out of plane $M_y$. A helical modulation with wavelength $\lambda \approx 85$ $\mathrm{\mu m}$ is clearly visible. Simulation values: $|\tilde{c}_2|\bar{n}_{3D}=320$ $\mathrm{Hz}$, $c_d\bar{n}_{3D}=0.8$ $\mathrm{Hz}$ and $q=100$ $\mathrm{Hz}$ (edges removed).}\label{tm}
  \end{center}
\end{figure}

For the $F_{\perp}$ state, a different modulation will appear.  In 
particular, since the state is XY-like (the rapid Larmor precession gives the same energy for all perpendicular spin directions), it can adopt a smoothly varying magnetization texture. 
The smoothest form will be a helix, with wave vector along the magnetic
field, see Fig.~\ref{phases}.  In other words, as shown in Fig.~\ref{tm}, the magnetization
will adopt a configuration like $M_x(z)=\sin{(k_z z)}$ and 
$M_y(z)=\sin{(k_z z+\frac{\pi}{2})}$ at any instant of time.
The kinetic energy of such a state goes as $k_z^2$, while the dipole
energy turns out to decrease as $k_z$ for small $k_z$; see
Appendix~\ref{app:helix} for details. 
\\At leading order in the dipole
strength, then,
\beq
\label{wv}
k_z^\perp \sim \frac{1}{\lambda_z^\perp} \sim \frac{ c_d \bar{n}_{3D}T}{\hslash^2/2m}
\eeq
with $\lambda_z^\perp$ the wavelength of the helical modulation, for a derivation see Appendix C. In $^{87}$Rb with experimentally
accessibly densities the wavelength is approximately
$80\mu\mathrm{m}$ and should be observable.  Note that the scales
for the two textures are related by
$d^\| \sim \lambda_z^\perp (L/\xi_S)$.

Since the modulations of $F_{\|}$ and $F_{\perp}$ decrease
the total energy of those states, their regions of the phase 
diagram, Fig.~\ref{cdpd}, will be larger than predicted in the 
previous subsection.  However, the dipole strength must be large 
to introduce domains into the $F_{\|}$ state; and the energy gain
in a helical texture relative to a uniform $F_{\perp}$ is small; 
so the phase boundaries will not change significantly at weak or moderate dipole strengths when we take
these textures into account.

\section{Numerical Results}
We investigate numerically the ground state phase
diagram of a spin-1 condensate in external fields that give rise to a
quadratic Zeeman shift and Larmor precession. The Metropolis
algorithm~\cite{metropolis} allows us to efficiently locate minima of
a given energy functional.  We discretize the system on a lattice, 
and for the fundamental move we draw random deviations in the six
real components of the field $\Psi$ from a normal distribution at a
lattice site.  The initial state is similarly generated from random
normally distributed variables.

A wide variety of simulation parameters ($N$, $a$, $\sigma_y$, $T_{MC}$, $T_{MF}^c$, $\mu$, $c_0$, see below), for example $1\times1<N<50\times50$, have been used to investigate
the phase diagram($c_2$, $q$, $c_d$). Energies have been calculated in Hz and the lengths
have been inserted in $\mu\mathrm{m}$. Unless otherwise noted,
numerical results presented here use lattice constant
$a=4\,\mu\mathrm{m}$, thickness $\sigma_y=2\,\mu\mathrm{m}$, and a system
size of $N=30\times30$ plaquettes.  We also add a chemical
potential to the energy, $\mu =1202\,\mathrm{Hz \mu m^{-2}}$, in order to reproduce
the experimental density for $c_0=1.9kHz$.  Finally, we set $T_{MC}=23\,\mathrm{nK}$ in
the Metropolis weight $e^{-\ev{H}/kT_{MC}}$, which strikes a good
 balance
between reducing fluctuations and achieving convergence in a reasonable 
computation time and use a critical mean field temperature $T_{MF}^c=100T_{MC}$.


The phase diagram we have mapped out numerically agrees well with the 
results presented so far.  In particular, we have confirmed
that the ferromagnetic states develop modulations governed by the
strength of the dipole interaction.

The algorithm described above tends to get trapped in local energy minima with varying densities of domain walls in the $F_{\|}$ region of the phase diagram. We can, however, locate the global minimum fairly confidently by starting the system in a variety of modulated states (striped or checkerboard) and comparing the final energies. The existence of metastable states as a consequence of dipolar interactions has been discussed before for spinor condensates in an optical lattice~\cite{dipolelattice}.  We have not observed any tendencies for the simulation in the $F_{\perp}/P_{\|}$ region of the phase diagram to be trapped in a local energy minima, regardless of the initial configuration. This is as expected, since any possible local ground state configuration (Eq.~\eqref{Fperppsi}) can smoothly turn into another, unlike in the $F_{\|}$ case (Eq.~\eqref{spfpa}). This symmetry between the two transverse components of the magnetization is present in the Hamiltonian without the dipole interaction, removed by the dipole interaction, and finally restored by the rapid Larmor precession.  However, even if the relaxational dynamics of the Metropolis algorithm used here does not apparently get trapped in a local minimum in this phase, the actual dynamics of the experimental system is primarily precessional rather than relaxational, which could lead to metastable states.


\subsection{Domain walls in $F_{\|}$}
\begin{figure}[htbp]
\begin{center}
    \subfigure{\includegraphics[width=81mm]{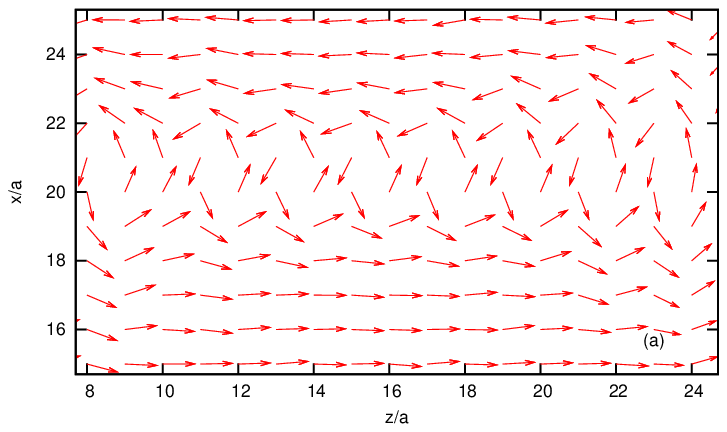}}\\
        \subfigure{\includegraphics[width=81mm]{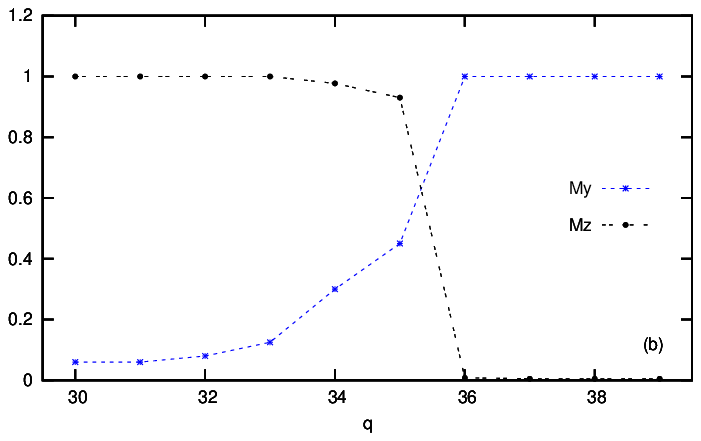}}
    \caption{(Color online) Transition to $F_{\perp}/P_{\|}$ from $F_{\|}$. (a) For $q$ slightly smaller than $q_{c2}$, large Mermin-Ho vortices appear between the stripes (plaquette size $a=4$ $\mathrm{\mu m}$). $M_z(x,z)$ is plotted on the horizontal axis, $M_{x}(x,z)$ on the vertical axis and $M_{y}(x,z)\approx 0$ for the whole region shown at this instant.  (b) Consequently, the maximum value of the Fourier transform of the magnetization out of plane $M_{x}(k_z^{max})$, see Eq.~\eqref{magfour}, increase before the phase transition.  
    Simulation variables: $|\tilde{c}_2|\bar{n}_{3D}=450$ $\mathrm{Hz}$, $c_d\bar{n}_{3D}=7.2$ $\mathrm{Hz}$ and $q=35$ $\mathrm{Hz}$ (a), $q=30-39$ $\mathrm{Hz}$ (b).}
    \label{Vor}
\end{center}
\end{figure}

Near the transition $q_{c2}(c_2,c_d)$, Eq.~\eqref{qc2}, magnetization vortices
with unit spin winding develop all the way along all domain walls, see Fig.~\ref{Vor}. The vortices are alternating elliptical and hyperbolic Mermin-Ho vortices, with ferromagnetic cores.~\cite{Ho,ZhangHo} The density of vortices increases with increasing dipole interaction, i.e. more domain walls appear and the longitudinal length of each vortex decreases. The transverse length of the vortices increases with increasing quadratic Zeeman strength up to the transition line,   
which can be seen in the Fourier transform of the magnetization
\beq
\label{magfour}
M_z(k_x)=\sum_{r,s}e^{-irk_x}M_z(r,s),
\eeq
as a rise in $M_y(k_z^{max})$; see Fig.~\ref{Vor} on
the $F^\|$ side of the transition.  The transition at $q_{c2}$  
itself remains sharp, and no vortices are observed for $q>q_{c2}$.  At a given instant in time does the perpendicular magnetization in all vortices in a domain boundary point in a specific direction.  The correlations between the direction of the transverse magnetization of vortices in different domain walls are however weaker. 


\subsection{Boundaries and trapping potential}
\begin{figure}[htbp]
  \begin{center}
    \includegraphics[width=81mm]{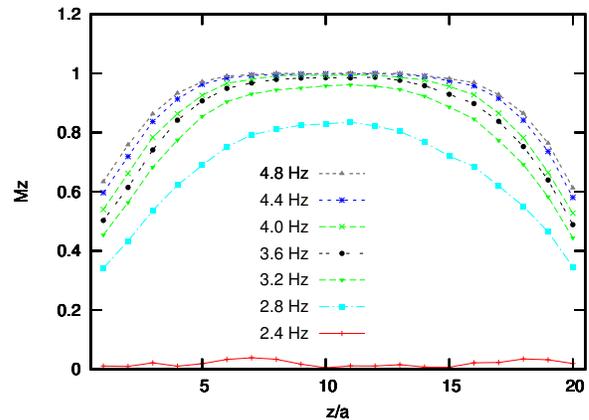}
    \caption{(Color online) Transition to $P_{\perp}$ from $F_{\|}$. The parallel magnetization $M_z(z/a)$ is plotted for different values of $|\tilde{c}_2|\bar{n}_{3D}=2.4 - 4.8$ $\mathrm{Hz}$ as a function of z/a. The magnetization is lowered at the boundaries around the transition point for a finite system. Simulation values: $N=20\times20$, $c_d\bar{n}_{3D}=5.7$ $\mathrm{Hz}$ and $q=-4$ $\mathrm{Hz}$.}
    \label{Transm}
  \end{center}
\end{figure}

\begin{figure}[htbp]
  \begin{center}
    \includegraphics[width=85mm]{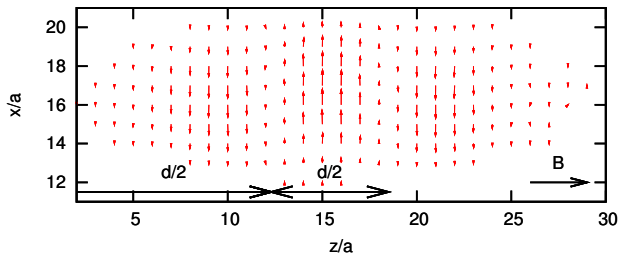}
    \caption{(Color online) Simulation of a helical modulated magnetized condensate in an elliptical trap.  $\bar{n}_{3D}(x,z)M_z(x,z)$ is plotted on the horizontal axis, $\bar{n}_{3D}(x,z)M_{x}(x,z)$ on the vertical axis, and $M_{y}(x,z)$ is a quarter of a wavelength ahead of $M_{x}(x,z)$ as in Fig.~\ref{tm}, but is not shown.  The wavelength $\lambda(z)$ of the helical modulation increases with decreasing density along the longitudinal axis.  The distance between two neighbouring nodes is shown; the node to the left of them is outside the graph.  Simulation parameters: $v_z=1$, $v_x=10$, $|c_2|\bar{n}_{3D}=540$ $\mathrm{Hz}$, $c_d\bar{n}_{3D}=1.6$ $\mathrm{Hz}$ and $q=120$ $\mathrm{Hz}$.}
    \label{trapw}
  \end{center}
\end{figure}

Finite size effects and the details of the trapping potential seem to
have little impact on our results.  The only finite size effect observed with hard-wall boundaries 
is a decrease in magnetization at the $z=\pm L/2$ boundaries in 
the transition from $F_{\|}$ to $P_{\perp}$, as shown in 
Fig.~\ref{Transm}. 
The approximative location of this transition line from the analytical calculation, Eq.~(\ref{FtP}), is $|\tilde{c}_{2c}|\bar{n}_{3D}=3.4\mathrm{Hz}$. \\
\\
We have also carried out simulations with an elliptical trap potential 
of the form $U(\bf{x})=U(v_z(\frac{z}{a})^2+v_x(\frac{x}{a})^2)$, typically
with $U=625$ $\mathrm{Hz \mu m^{-2}}$ and $v_z,v_x=1-10$ to more closely model experimental
conditions.~\cite{Sadler,Veng1,Veng2}  These simulations have shown no 
effect other than a decrease in the density and thereby related effects, as in the original paper of Ho on spinor condensates in optical traps.~\cite{Ho}  For example, the wavelength of the helical modulation in $F_{\perp}/P_{\|}$ is inversely proportional to the density, see Fig.~\ref{trapw} which shows a change in wavelength through the condensate as the density changes.  In particular, we have not seen the effect reported by Vengalattore
\etal~\cite{Veng2} in which the modulation wave
vector is not aligned with the applied magnetic field but is instead
influenced by the orientation of the trap. 

\section{Discussion}
We have mapped out the complete phase diagram for the model we have 
considered.  Although the region occupied by the phase 
$F_{\perp}/P_{\|}$ moves and shrinks with the introduction of the dipole 
interaction, we find that it remains accessible at physical values of 
$|c_2|$ and $c_d$ in $^{87}$Rb, for some values of the quadratic Zeeman 
shift $q$. Hence, by tuning $q$ for $^{87}$Rb appropriately, the 
three phases $F_{\|}$, $F_{\perp}/P_{\|}$ and $P_{\|}$ should be 
observable in experiments.  We also find that a spatial modulations 
should be seen in at least the second of those phases.

There are some disagreements between our result and other results 
obtained theoretically and more important experimentally.  The length scale in the experiment is smaller than the pitch of the helical modulation we describe above by a factor 10, roughly, for typical parameters.  Cherng and Demler~\cite{CherngDemler} find a dynamical instability at a scale nearer that seen in experiment.  That picture would suggest that even if the phase diagram obtained here describes the system at long times,
the experimental system might instead reach a long-lived metastable state.
As explained in section V above, while we do see metastable states in some
parts of the phase diagram, we do not see metastable checkerboard states in
the region probed by current experiments, but this could be because the Metropolis
dynamics of our simulation is not the actual dynamics of the condensate, even if
their thermodynamics are the same.

One challenge for this dynamical scenario is that in experiments, an imposed helical configuration with pitch $\lambda=50-150\mu \mathrm{m}$~\cite{Veng1} 
quickly evolves into a state modulated at a smaller scale, again roughly 
ten times smaller than the stable, or at least metastable, supersolid 
state we predict.~\cite{Veng1,Veng2}  This suggests that effects we 
have not taken into account prevent the current experimental system 
from finding this minimum.  As an example, it is known that the dipole 
interaction makes the Larmor precession unstable, according to 
Lamacraft~\cite{Lamacraft}; as a result, the Larmor-averaged energy that is
the main focus of the present work might not be an accurate description for long times.

In order to observe the predicted supersolid clearly, our results 
suggest that the key is to suppress this Larmor instability while at the 
same time preserving the conservation of total magnetization in the 
field direction.  The Larmor instability~\cite{Lamacraft} grows 
exponentially from thermal excitation of an initial perturbation at the 
Larmor frequency $\omega_L$.  Hence the time scale to reach a fixed 
final size of the instability is proportional to $\hbar \omega_L / (k_B 
T)$ and can be increased either by increasing the magnetic field or 
decreasing the temperature.  At the same time, an experiment should be 
designed to preserve the magnetization along the field direction for as 
long as possible, which requires a high degree of trap uniformity.  One motivation
for continued exploration of this system is that our results show that the
Larmor-averaged system does have a supersolid ground state for a wide
range of parameters.

{\it Note added}: As this work was being prepared for submission, two e-prints appeared investigating the same experiment by slightly different approaches.~\cite{ZhangHo,kawaguchi}  The first, by J.~Zhang and T.-L.~Ho, also investigates the static properties of $^{87}$Rb using a deterministic numerical method and also gets the $F_{\|}$ state and a modulated $F_{\bot}$ state.  The main difference between their results and ours appears to be that they find a stripe phase rather than a helix for the phase with spins perpendicular to the applied magnetic field.  They find arrays of elliptical and hyperbolic Mermin-Ho vortices, as a meta-stable dynamical state, between the stripes for the $F_{\|}$ state for all q. However, they are smaller than the spin healing length and hence unobservable in our simulation, although we do see them close to the transition to the $F_{\bot}/P_{\|}$ state.  The second, by Y.~Kawaguchi {\it et al.}, finds a doubly periodic (checkerboard) spin pattern as a long-lived intermediate state through a combination of mean-field theory and numerical simulation of precession-averaged equations of motion. By adding energy dissipation to the dynamics, they reach a stationary state similar to ours.

\appendix
\section{The dipole term}\label{app:dipole}
The dipolar energy of a magnetized fluid with magnetization 
$\bm{\mathcal{M}}(\bf{x})$ is
\begin{multline}
\nonumber \frac{\mu_0}{8\pi} \int \! d\bf{x}d\bf{x}'
  \left[ \frac{ \bm{\mathcal{M}}\cdot\bm{\mathcal{M}}'
  - 3(\bm{\mathcal{M}}\cdot\hat{\bf{r}}) (\bm{\mathcal{M}}'\cdot\hat{\bf{r}}) }{r^3}
  \right. \\
  \left. - \frac{8\pi}{3} \mathcal{M}^2 \delta^{(3)}(\bf{r})
  \right],
\end{multline}
where $\bf{r} = \bf{x}-\bf{x}'$ and $\bm{\mathcal{M}}' = \bm{\mathcal{M}}(\bf{x}')$.  The last
term, or ``s-wave'' part, contributes to the contact interaction $c_2$ in
the BEC Hamiltonian, and so should not be treated independently.  In this
paper we take the first, ``d-wave'' part to be the full dipolar interaction.
This can, in turn, be decomposed into a ``Coulomb'' part that is
positive semidefinite, and hence convenient for numerical work that searches
for energy minima, and a contact part, as in Eq.~\eqref{dipham}.

For both analytical and numerical work we need the dimensionally
reduced form of the Coulomb part expressed in a rotating
frame.  Ignoring the contact term in $H_{dip}$ and performing two
partial integrations we find
\begin{widetext}
\begin{align}
        \label{al:Ed}
\nonumber
E_{dip}^C &= \frac{c_d}{2}\int d^3xd^3x' 
\frac{\bm{\nabla}\cdot (n_{3D}\bm{M}(\bf{x})) \bm{\nabla}'\cdot(n_{3D}\bm{M}(\bf{x}'))}{|\bf{x}-\bf{x}'|} 
= \frac{c_d}{2}\int d^2xd^2x' \sigma(x,z) \sigma'(x',z')
\int dydy' \frac{\rho(y) \rho(y')}{|\bf{x}-\bf{x}'|}+\\
&\frac{c_d}{2}\int d^2x d^2x' n_{2D}M_y(x,z)n_{2D}M_y(x',z')
\int dydy'
\frac{[\partial_y \rho(y)][\partial_{y'} \rho(y')]}{|\bf{x}-\bf{x}'|}
\end{align}
\end{widetext}
where $\sigma(x,z) \equiv \partial_x (n_{2D}M_x(x,z)) + \partial_z (n_{2D}M_z(x,z))$ is
an effective surface charge density. The density $n_{2D}$ has only a $(x,z)$ dependence for a nonzero trapping potential $U(x,z)$.
The integrals over $y$ can be performed explicitly for either
Gaussian or boxcar profiles $\rho$; we choose the Gaussian form for
the purposes of numerics.  Then
\begin{align}
\rho(y) \rho(y') &= \frac{2}{\pi \sigma_y^2} e^{-(y_+^2 + y_-^2)/\sigma_y^2} \notag \\
[\partial_y \rho(y)][\partial_{y'} \rho(y')]
&= \frac{8(y_+^2 - y_-^2)}{\pi \sigma_y^6} e^{-(y_+^2 + y_-^2)/\sigma_y^2}
\end{align}
with $y_{\pm} = y \pm y'$.  The integrals over $y_+$ are simple, and
the integrals over $y_-$ can be put in terms of special functions with help of the identities $\int 
dx\frac{e^{-x^2}}{\sqrt{c^2+x^2}}=e^{\frac{c^2}{2}}K_0(\frac{c^2}{2})$ and
$\int dx \frac{x^2e^{-x^2}}{\sqrt{c^2+x^2}}
=\frac{\sqrt{\pi}}{2}U(\frac{1}{2},0,c^2)$.  Here $K_0$ is a
modified Bessel function and $U$ is a confluent hypergeometric
function.

For the numerics, discretize the remaining integrals as follows.
Divide the 2-dimensional area into rectangular plaquettes and set
the density $n_{2D}$ and magnetization $\bm{M}$ constant on each plaquette,
\beq\label{al:Ed2}
\bm{M}(x,z) \rightarrow \bm{M}(a (r+\frac{1}{2}),a (s+\frac{1}{2})),
\eeq
where $a$ is the lattice constant and $r,s$ are integers.
Then do several variable substitutions.  Going to variables $x_\pm$
and $z_\pm$ and scaling the coordinates by $a$ allows us to replace
\beq
\int \!d^2 x d^2 x' \rightarrow \!
\int_{p-1}^{p+1} \!\!dx_- \! \int_{q-1}^{q+1} \!\!dz_-
(1-|x_- -p|)(1-|z_- -q|)
\eeq
since the integrands depend only on $x_-,z_-$.  Here $p=r'-r$ and
$q=s'-s$.

The integrals can then be computed numerically for $0\le p,q<\sqrt{N}$.
The final step is to time-average the fields to take into account the
rapid Larmor precession. This effectively means replacing
\begin{equation}
\begin{split}
\sigma(p,q) \sigma(p',q') &\rightarrow
\partial_z(n_{2D} M_z(p,q)) \partial_{z'}(n_{2D} M_z(p',q')) \\
&+ \frac{1}{2} \partial_x(n_{2D} M_x(p,q)) \partial_{x'}(n_{2D} 
M_x(p',q') )\\
&+ \frac{1}{2} \partial_x(n_{2D} M_y(p,q)) \partial_{x'}(n_{2D} M_y(p',q')
)
\end{split}
\end{equation}
and
\beq
\begin{split}
M_y(p,q)M_y(p',q') &\rightarrow
\frac{1}{2} M_x(p,q)M_x(p',q') \\
&+ \frac{1}{2} M_y(p,q)M_y(p',q')
\end{split}
\eeq
in Eq.~\eqref{al:Ed}, since the transverse components rotate into
each other but the longitudinal component is unaffected.

\section{Helical modulation}\label{app:helix}
We can obtain a simple estimate of the wavelength of the transverse 
helical state
to leading order in the strength of the dipole coupling by assuming a fully
polarized time evolving state $\psi^F_\perp(0,k_zz-\gamma B_0 t)$, 
Eq.~\eqref{Fperppsi}, with magnetization
\begin{equation}
 M_x + i M_y = n_{2D}\rho(y) e^{i(k_zz-\gamma B_0 t)}.
\end{equation}
Fourier transforming the kinetic and the dipole energy term, keeping only
contributions that scale with the area of the two-dimensional system, the
(areal) energy density of this state is
\begin{equation}
 \frac{\mathrm{energy}}{\mathrm{area}} = n_{2D} \frac{\hslash^2}{2m} 
  \frac{k_z^2}{2} + \frac{c_d}{2}\frac{n_{2D}^2}{2} \int \!\frac{dk_y}{2\pi}\,
	\frac{4\pi}{k_y^2+k_z^2}k_y^2 |\tilde{\rho}(k_y)|^2
\end{equation}
plus $k_z$-independent terms.  In the kinetic term, there is a factor 
of $1/2$ because only half the atoms are in the $m_z = \pm 1$ states that 
carry kinetic energy.  In the dipole term, the only extensive contribution to
the energy comes from the out-of-plane component $M_y$, which gives a factor
$1/2$ there as well. Notice also that the time dependence is gone. With
$k_y^2/(k_y^2 + k_z^2) = 1- k_z^2/(k_y^2 + k_z^2)$, the
relevant terms are
\begin{equation}
 n_{2D} \frac{\hslash^2}{2m} 
  \frac{k_z^2}{2} - \frac{c_d}{2}\frac{n_{2D}^2}{2} |k_z|\int \!\frac{du}{2\pi}\,
	\frac{4\pi}{1+u^2} |\tilde{\rho}(|k_z|u)|^2,
\end{equation}
and to lowest order in $k_z$ we just need 
$\tilde{\rho}(0) = \int \!dy\, \rho(y) = 1$ to arrive at
\begin{equation}
 n_{2D} \frac{\hslash^2}{2m} \frac{k_z^2}{2} - \pi \frac{c_d}{2}n_{2D}^2 |k_z|,
\end{equation}
which takes its minimum at
\begin{equation}
 k_z = \pm \frac{\pi}{2} \frac{n_{2D}c_d}{\hslash^2/2m}.
\end{equation}

\section{Dipole energy at uniform magnetization}\label{app:Dipener}
For a uniform condensate with maximal magnetization, aligned
parallel to the magnetic field, the only contribution to the dipole
energy comes from from the edges at $z=\pm L/2$. The second term
in Eq.~(\ref{al:Ed}) does not contribute and only the edges of the
first
\begin{multline}\label{DEu}
E_{dip}^{C} = \frac{c_d n_{2D}^2}{2}
\int_{-W/2}^{W/2} dx dx' \int_{-\infty}^{\infty} dydy' \rho(y) \rho(y')
\\
\times 2 \left[ \frac{1}{\sqrt{x_{-}^2+y_{-}^2}}
-\frac{1}{\sqrt{x_{-}^2+y_{-}^2+L^2}} \right]
\end{multline}
In the limit $L \gg W \gg T$, the leading contribution
to the energy comes solely from the first term, which describes
the self energy of two quasi-one-dimensional lines of charge.  Indeed,
it becomes just
\begin{align}
\label{DEu2}
E_{dip}^{C} &= 2 c_d n_{2D}^2 \int^{W}\! dx_- (W-x_-)/x_-
  \notag \\
&= 2 c_d n_{2D}^2 W \ln W/T + O(W)
\end{align}
asymptotically, where the lower cutoff $T$ has been chosen for
convenience.

The energy for the uniform out-of-plane configuration is
\begin{equation}
  E_{dip}^{C} = \frac{c_d}{2} n_{2D}^2 \int d^3x d^3x'
\frac{[\partial_y \rho(y)][\partial_{y'} \rho(y')]}{|\bf{x}-\bf{x}'|}.
\end{equation}
Since there will be a term extensive in the planar size, it is simplest
to ignore the effects of boundaries and work with a surface energy
density
\begin{align}
  \eta &= \frac{c_d}{2} n_{2D}^2 2\pi \int \! dy dy' \int_0^R \!dr \,
  \frac{r \, [\partial_y \rho(y)][\partial_{y'} \rho(y')] 
}{\sqrt{r^2+y_-^2}}
  \notag \\
  &= 2\pi c_d n_{2D}^2 \int dy dy' \rho(y) \rho(y') \delta(y-y') + O(1/R)
  \notag \\
  &= 2\pi c_d n_{2D}^2 \frac{1}{T}
\end{align}
after integrating over the radial coordinate $r$ followed by partial
integration in $y$ and $y'$.

\acknowledgments

The authors thank Subroto Mukerjee, Dan Stamper-Kurn, Mukund Vengalattore, Kater Murch, Jennie Guzman, Andre Wenz, Ari Turner, and Ashvin Vishwanath for useful comments and acknowledge support from ARO through the OLE program (J.~K., J.~E.~M), Knut and Alice Wallenberg foundation (J.~K.) and WIN (A.~E.).

\end{document}